\documentclass[journal,12pt,onecolumn,draftclsnofoot,]{IEEEtran}

\usepackage{algorithm}
\usepackage{algpseudocode}
\usepackage{graphicx}
\usepackage{subcaption}
\usepackage{booktabs}
\usepackage{enumitem}
\usepackage{amssymb}
\usepackage{multirow}
\usepackage{cite}
\usepackage{color}
\usepackage{color}
\usepackage{hyperref}

\usepackage[cmex10]{amsmath}

\begin{document}

%

\title{Hyperspectral unmixing with spectral variability using adaptive bundles \\and double sparsity}

%

\author{Tatsumi~Uezato, Mathieu Fauvel,~\IEEEmembership{Senior Member,~IEEE,}
        \\and Nicolas Dobigeon,~\IEEEmembership{Senior Member,~IEEE}
\thanks{Part of this work has been funded by EU FP7 through the ERANETMED
JC-WATER program, MapInvPlnt Project ANR-15-NMED-
0002-02 and by the MUESLI IDEX ATS  project, Toulouse INP.}
\thanks{T. Uezato and N. Dobigeon are with University of Toulouse, IRIT/INP-ENSEEIHT, CNRS, 2 rue Charles Camichel, BP 7122, 31071 Toulouse Cedex 7, France (e-mail: \{Tatsumi.Uezato, Nicolas.Dobigeon\}@enseeiht.fr).}
\thanks{M. Fauvel is with University of Toulouse, INRA, DYNAFOR, BP 32607, Auzeville-Tolosane, 31326 Castanet Tolosan, France (e-mail: mathieu.fauvel@ensat.fr).}
 }

\maketitle

\begin{abstract}
Spectral variability is one of the major issue when conducting hyperspectral unmixing. Within a given image composed of some elementary materials (herein referred to as endmember classes), the spectral signature characterizing these classes may spatially vary due to intrinsic component fluctuations or external factors (illumination). These redundant multiple endmember spectra within each class adversely affect the performance of unmixing methods. This paper proposes a mixing model that explicitly incorporates a hierarchical structure of redundant multiple spectra representing each class. The proposed method is designed to promote sparsity on the selection of both spectra and classes within each pixel. The resulting unmixing algorithm is able to adaptively recover several bundles of endmember spectra associated with each class and robustly estimate abundances. In addition, its flexibility allows a variable number of classes to be present within each pixel of the hyperspectral image to be unmixed. The proposed method is compared with other state-of-the-art unmixing methods that incorporate sparsity using both simulated and real hyperspectral data. The results show that the proposed method can successfully determine the variable number of classes present within each class and estimate the corresponding class abundances.
\end{abstract}

\newpage

\IEEEpeerreviewmaketitle


\section{Introduction}

\IEEEPARstart{H}{yperspectral} analysis has received an increasing attention because of its high spectral resolution, which enables a variety of objects to be identified and classified \cite{Plaza2009}. \textit{Mixed pixels} caused by the presence of multiple objects within a single pixel adversely affect the performance of hyperspectral analysis~\cite{Kesha2002}. To address this problem, a wide variety of spectral unmixing methods have been developed over the last decades~\cite{Biouc2012,Dobig2014,Heyle2014,Uezat2017}. Spectral unmixing methods aim at decomposing a mixed spectrum into a collection of reference spectra (known as \textit{endmembers}) characterizing the macroscopic materials present in the scene, and their respective proportions (known as \textit{abundances}) in each image pixel~\cite{Kesha2002}. Despite the large number of developed spectral unmixing methods, there are still major challenges for accurate estimates of endmember signatures and abundances~\cite{Biouc2012}. Among these challenges, endmember variability may lead to large amounts of errors in abundance estimates~\cite{Zare2014}. It results from the fact that each endmember can rarely be represented by a unique spectral signature. Conversely, it is subject to so-called spectral variability, e.g., caused by variations in the acquisition process, the intensity of illumination or other physical characteristics of the materials~\cite{Murph2012,Uezat2016a}. Taking this endmember variability into account during the spectral unmixing process is one of keys for successful application of spectral unmixing~\cite{Somer2011}.

The methods that incorporate endmember variability can be categorized into two main approaches (see \cite{Somer2011,Zare2014,Drume2016b} for recent overviews). The first approach relies on the definition of a set of multiple spectral signatures, referred to as endmember bundles, to characterize each endmember class. Endmember bundles can be collected from field campaign  or can be extracted from data itself using endmember bundle extraction methods~\cite{Bates2000,Somer2012,Uezat2016a}. The advantage of this approach is that the method can use \textit{a priori} information representing endmember variability. Endmember bundles can be validated by experts in order to provide accurate representation about endmember variability~\cite{Bates2000}. Although traditional methods incorporating endmember bundles (e.g.~\cite{Rober1998}) are known to be computationally expensive, more  efficient methods have been recently developed and have shown great potential~\cite{Vegan2014,Goena2013,Uezat2016b,Meyer2016}. However, as pointed out in~\cite{Uezat2016}, it is unlikely that the endmember bundles completely represent endmember variability present in an image. Such incomplete endmember bundles may lead to poor estimates of abundances. The second approach uses physical or statistical descriptions of the endmember variability. More precisely, these methods describes the endmember variability thanks to a statistical distribution~\cite{Halim2015} or by incorporating additional variability terms in the mixing model~\cite{Drume2016,Thouv2016,Hong2017}. The advantage of this approach results from the adaptive learning of the endmember variability. Indeed, state-of-the-art methods such as those recently introduced in \cite{Drume2016,Thouv2016,Hong2017} enable endmember spectra to spatially vary within each pixel in order to describe endmember variability. This is important since the endmember spectra to be used for the abundance estimation can be different between pixels. However, estimating endmember variability without \textit{a priori} knowledge is a challenging task, especially when large amounts of endmember variability are present in an image. In addition, the statistical distribution or additional terms used in these methods may be overly simplified to represent endmember variability.

Both approaches demonstrate benefits and drawbacks. A natural question arises: is it possible to combine the strong advantages of both approaches to robustly represent endmember variability? This paper addresses the question and introduces a novel spectral unmixing method that bridges the gap between the aforementioned two approaches with the help of a double sparsity-based method inspired by \cite{Rubin2010}. Specifically, the proposed method aims at adaptively recovering endmember spectra within each pixel to describe endmember variability while incorporating available \textit{a priori} information. The proposed method is closely related to the existing methods. Thus, the main contributions of this paper are threefold: 1) to propose a novel spectral unmixing method that incorporates endmember bundles and generates adaptive endmember spectra within each pixel, 2) to give a systematic review of related work and show the relationship between the proposed method and existing methods, 3) to provide comparison between the proposed method and other sparsity-based methods.

This paper is organized as follows. Section \ref{section2} describes related works and existing methods, while highlighting their inherent drawbacks. In Section \ref{section3}, a novel mixing model that incorporates endmember variability is proposed and its relationships with existing methods are discussed. Section \ref{sec:algo} introduces an associated unmixing algorithm designed to recover the endmember classes, adaptive bundles and abundances. Section \ref{section4} and Section \ref{section5} show experimental results obtained from simulated data and real hyperspectral images. Finally, conclusions are drawn in Section \ref{section6}.

\section{Related works and issues raised by existing methods}\label{section2}
\subsection{Conventional (variability-free) linear mixing model}
Let $\mathbf{y}_i \in \mathbb{R}^{L \times 1}$ denote the $L$-spectrum measured at the $i$th pixel of an hyperspectral image. According to the linear mixing model (LMM), the observed spectrum of the $i$th pixel $\mathbf{y}_i$ is approximated by a weighted linear combination of endmember spectra and abundance fractions
\begin{eqnarray}
\label{eq:LMM}
  \mathbf{y}_i=\mathbf{M}\mathbf{a}_i+\mathbf{n}_i
\end{eqnarray}
where $\mathbf{M} \in \mathbb{R}^{L \times K}$ is the matrix of the spectral signatures associated with the $K$ endmember classes, $\mathbf{a}_i = \left[a_{1i},\ldots,a_{Ki}\right]^T \in \mathbb{R}^{K \times 1}$ is the abundance fractions of the pixel and $\mathbf{n}_i \in \mathbb{R}^{L \times 1}$ represents noise and modeling error. LMM is generally accompanied by abundance non-negativity constraint (ANC) and the abundance sum-to-one constraint (ASC)
\begin{eqnarray}
    \forall k, \forall i, \   a_{ki}\geq 0,\quad \text{and} \quad  \forall i,\  \sum_{k=1}^{K}a_{ki}=1
\end{eqnarray}
where $a_{ki}$ is the abundance fraction of the $k$th class in the $i$th pixel. This model implicitly relies on two assumptions: \emph{i}) each endmember class is described by a unique spectrum and \emph{ii}) the endmember matrix $\mathbf{M}$ is fixed and commonly used to unmix all the pixels of a given image. In other words, LMM does not account for spectral variability. However, as discussed previously, this is likely unrealistic because spectral variability is naturally observed in hyperspectral images, e.g., because of variations in illumination or  physical intrinsic characteristics of materials \cite{Uezat2016}.

\subsection{Linear mixing models incorporating endmember bundles}
Multiple endmember spectral mixture analysis (MESMA) \cite{Rober1998} allows the variability of the endmember spectrum representative of each class and a varying number of endmember classes present within each pixel. Although MESMA has been widely used for a variety of applications \cite{Rober1998,Denni2003}, it owns several major limitations: \emph{i}) it is highly computationally expensive because MESMA needs to test a large number of combinations of endmember spectra \cite{Tits2013}, \emph{ii}) MESMA tends to select an overestimated number of endmember classes because it uses the reconstruction error to select the appropriate combination of endmember spectra \cite{Demar2012} and \emph{iii}) the performance of MESMA may significantly decrease when endmember spectra (or \emph{bundles}) within each class do not completely represent the spectral variability \cite{Uezat2016}.

To overcome these limitations, recent works have proposed a new class of methods that incorporate all endmember bundles  defined as \cite{Goena2013,Vegan2014}
\begin{eqnarray}
\label{eq:bundles}
  \mathbf{E}=\begin{bmatrix}\ \mathbf{E}_1 \ | \ \mathbf{E}_2\  |\cdots|\  \mathbf{E}_K\ \end{bmatrix}
\end{eqnarray}
where $\mathbf{E}_k \in \mathbb{R}^{L \times N_k}$ represents a set of endmember spectra (aka bundle) characterizing the $k$th class, $N_k$ is the number of endmember spectra in the $k$th class and $N$ is the total number of endmember spectra of all classes with $N= \sum_{k=1}^K N_k$. Generalizing LMM in \eqref{eq:LMM}, those methods firstly model a given observed pixel spectrum with respect to all spectra in endmember bundles and corresponding multiple abundances
\begin{eqnarray}\label{eq:4}
  \mathbf{y}_i=\mathbf{E}\mathbf{r}_i+\mathbf{n}_i
\end{eqnarray}
where $\mathbf{r}_i \in \mathbb{R}^{N \times 1}$ is multiple abundance fractions corresponding to each spectrum of the endmember bundles $\mathbf{E}$. As for LMM, ASC or ANC can also be imposed to $\mathbf{r}_i$. As a second step, multiple abundance fractions $\mathbf{r}_i$ are summed within each class to generate a single abundance fraction for each class
\begin{eqnarray}\label{summed}
  \mathbf{a}_i=\mathbf{G}^T\mathbf{r}_i
\end{eqnarray}
with
\begin{eqnarray}
  \mathbf{G} = \begin{bmatrix} \mathbf{1}_{N_1} & \mathbf{0}_{N_1} & \dotsb & \mathbf{0}_{N_1}
    \\  \mathbf{0}_{N_2} & \mathbf{1}_{N_2} & \mathbf{\dotsb} & \mathbf{0}_{N_2}
\\ \mathbf{\vdots} & \mathbf{\vdots} & \mathbf{\ddots} & \mathbf{\vdots}
\\ \mathbf{0}_{N_K} & \mathbf{0}_{N_K} & \mathbf{\dotsb} & \mathbf{1}_{N_K} \end{bmatrix}
\end{eqnarray}
where $\mathbf{1}_{N_k} \in \mathbb{R}^{N_k \times 1}$ is a column vector of ones and $\mathbf{0}_{N_k} \in \mathbb{R}^{N_k \times 1}$ represent a $N_k$-dimensional vector whose components are zeros. While these two steps are conducted separately in \cite{Goena2013,Vegan2014}, they can be also considered jointly  within a multi-task Gaussian process framework \cite{Uezat2016,Uezat2016b}. Even if these methods have been shown to be effective, a large number of endmember spectra within each class may be redundant. In such case, following a model selection inspiration, Veganzones \emph{et al.} introduce a complementary sparsity regularization on the multiple abundance vectors \cite{Vegan2014}
\begin{eqnarray}
\begin{aligned}
	& \min\limits_{\mathbf{r}_i}\frac{1}{2}\Vert\mathbf{E}\mathbf{r}_i-\mathbf{y}_i\Vert^2_2+\lambda_r\Vert\mathbf{r}_i\Vert_1 \\
    & \text{s.t. }   \forall i, \quad  \mathbf{r}_{i} \succeq 0
\end{aligned}
\end{eqnarray}
where $\succeq$ represents the element-wise comparison, $\Vert\cdot\Vert_2$ is the $\ell_2$-norm, $\Vert\cdot\Vert_1$ is the $\ell_1$-norm which is known to promote sparsity. Once the multiple abundance vector $\mathbf{r}_i$ has been estimated, it is normalized in order to reduce the effects of multiplicative factors and satisfy ASC. Following the same approach, further sparsity can be imposed using $\ell_p$-norm \cite{Sigur2014} or reweighted $\ell_1$-approaches \cite{Zheng2016,He2017}. Overall, this sparsity property allows the selection of a smaller number of endmember spectra. However, it may not lead to the selection of a smaller number of endmember classes. Conversely, to promote sparsity on the number of endmember classes, one strategy consists in formulating the unmixing problem through a sparse group lasso \cite{Iorda2011a}
\begin{eqnarray}
\begin{aligned}
	& \min\limits_{\mathbf{r}_i}\left\lbrace \frac{1}{2}\Vert\sum_{k=1}^{K}\mathbf{E}_k(\mathbf{g}_k\odot\mathbf{r}_i)-\mathbf{y}_i\Vert^2_2 \right. \\
    & \left. +\lambda_g\sum_{k=1}^{K}\Vert\mathbf{g}_k\odot\mathbf{r}_i\Vert_2 + \lambda_r\Vert\mathbf{r}_i\Vert_1 \right\rbrace\\
    & \text{s.t. }   \forall i, \quad  \mathbf{r}_{i} \succeq 0
\end{aligned}
\end{eqnarray}
where $\mathbf{g}_k$ is the $k$th column of $\mathbf{G}$, $\odot$ is the element-wise product and thus $\mathbf{g}_k\odot\mathbf{r}_i$ extracts the elements in $\mathbf{r}_i$ belonging to the $k$th class. This approach has the great advantage of promoting sparsity in both the number of endmember spectra and the number of endmember classes. Another strategy relies on the concept of ``social sparsity'' that can exploit the structure of endmember bundles more explicitly \cite{Meyer2016}. The method assumes that $\mathbf{r}_i$ can be partitioned into $K$ groups representing each endmember class, leading to the optimization problem
\begin{eqnarray}
\begin{aligned}
	& \min\limits_{\mathbf{r}_i} \left\lbrace \frac{1}{2}\Vert\mathbf{E}\mathbf{r}_i-\mathbf{y}_i\Vert^2_2+\lambda_r\left(\sum_{k=1}^{K}\Vert \mathbf{g}_k\odot\mathbf{r}_i\Vert_p^q\right)^{\frac{1}{q}} \right\rbrace\\
    & \text{s.t. }  \forall i,\forall n, \quad  \mathbf{r}_{i} \succeq 0, \quad \sum_{n=1}^{N}r_{ni}=1
\end{aligned}
\end{eqnarray}
where $\Vert\cdot\Vert_p$ is the $\ell_p$-norm and $r_{ni}$ is the $n$th multiple abundance fraction of the $i$th pixel.  Finally, abundances associated each endmember class can be obtained by summing the multiple abundances within each class as in \eqref{summed}. This method can be considered as a generalized model since, by adjusting the values of $(p,q)$, it boils down to the group lasso, the elitist lasso or the fractional case \cite{Meyer2016}.

However, all aforementioned sparsity-based methods still suffer from the following limitations:
\begin{enumerate}
  \item \emph{Physically unrealistic abundance fractions}: They explicitly generate unrealistic multiple abundances corresponding to each spectrum in endmember bundles.
  \item \emph{Lack of adaptability to describe endmember variability}: ASC imposed on $\mathbf{r}_i$ does not allow a consistant description of the endmembers within each pixel. In addition, it cannot capture adaptive and hierarchical structure of endmember spectra for each pixel.
\end{enumerate}
To overcome these two shorcomings, the present paper capitalizes on this abundant literature to design a new multiple endember mixing model introduced in the next section.

\section{Multiple endmember mixing models}\label{section3}
\subsection{MEMM}
The proposed model relies on 3 main ingredients, namely endmember bundles, bundling coefficients and abundances. According to this model, each endmember bundle is mixed to provide a suitable and adaptive endmember spectrum used to unmix a given pixel. The proposed multiple endmember mixing model (MEMM) is defined as
\begin{eqnarray}
\label{eq:MEMM_model}
  \mathbf{y}_i=\mathbf{E}\mathbf{B}_i\mathbf{a}_i+\mathbf{n}_i
\end{eqnarray}
where $\mathbf{B}_i \in \mathbb{R}^{N \times K}$ gathers so-called bundling coefficients of the $i$th pixel which decompose the endmember signatures according to the endmember bundles for the considered pixel. To enforce the bundle structure, the bundling coefficients $\mathbf{B}_i$ associated with the pixel is defined as the following block-diagonal matrix
\begin{eqnarray}
  \mathbf{B}_i = \begin{bmatrix} \mathbf{b}_{1i} & \mathbf{0}_{N_1} & \dotsb & \mathbf{0}_{N_1}
    \\  \mathbf{0}_{N_2} & \mathbf{b}_{2i} & \mathbf{\dotsb} & \mathbf{0}_{N_2}
\\ \mathbf{\vdots} & \mathbf{\vdots} & \mathbf{\ddots} & \mathbf{\vdots}
\\ \mathbf{0}_{N_K} & \mathbf{0}_{N_K} & \mathbf{\dotsb} & \mathbf{b}_{Ki} \end{bmatrix}
\end{eqnarray}
where $\mathbf{b}_{ki} \in \mathbb{R}^{N_k \times 1}$ is the bundling coefficients for the $k$th class at the $i$th pixel. Each bundling coefficient must be nonnegative and the bundling vector $\mathbf{b}_{ki}$ is expected to be sparse. Indeed multiple endmember spectra within each class are usually redundant and only a few endmember spectra within each class should be enough to unmix a pixel. This property can be induced by considering the following bundling constraints
\begin{equation}
\label{eq:B_constraints}
    \forall i, \ \mathbf{B}_i \succeq 0 \quad \text{and} \quad  \Vert\mathbf{B}_i\Vert_0=\sum_{k=1}^{K}\Vert\mathbf{b}_{ki}\Vert_0 \leq s
\end{equation}
where $\Vert\cdot\Vert_0$ is the $\ell_0$-norm that counts the number of nonzero elements and $s$ is the maximum number of nonzero elements in $\mathbf{B}_i$, i.e., the maximum number of endmembers to be used within each class to describe the pixel. The abundance non-negativity constraint (ANC) and the abundance sum-to-one constraint (ASC) are usually imposed. In addition, in this work, complementary sparsity is imposed on each abundance vector, i.e.,
\begin{equation}
\label{eq:A_constraints_MEMM}
  \forall k,\forall i,\  a_{ki}\geq 0,\quad \text{and} \quad \forall i,\ \sum_{k=1}^{K}a_{ki}=1, \ \ \Vert\mathbf{a}_i\Vert_0 \leq v
\end{equation}
where $v$ is the number of endmember classes to be used to decompose the image pixel.

\subsection{MEMMs}\label{subsec:MEMMs}
The sparsity constraint \eqref{eq:B_constraints} applied to $\mathbf{B}$ can be slightly modified to obtain another meaningful set of constraints
\begin{equation}
\label{eq:B_constraints_MEMMs}
    \forall i,\ \mathbf{B}_i \succeq 0 \quad \text{and} \quad \forall k, \forall i,\ \Vert\mathbf{b}_{ki}\Vert_0 \leq 1.
\end{equation}
The resulting model, referred to as MEMM$_s$ in what follows, is designed to generate at most one scaled endmember spectrum for each class.


\subsection{Relationships between MEMM and existing models}
\subsubsection{MEMM$_s$ and MESMA}
When the sparsity constraint on the abundances $\Vert\mathbf{a}_i\Vert_0 \leq v$ in \eqref{eq:A_constraints_MEMM} is not considered, the optimization problem associated with the MEMM$_s$ model described in paragraph \ref{subsec:MEMMs} is equivalent to MESMA and sparse MESMA \cite{Chen2016}. Unlike MESMA that considers the reconstruction error to determine the optimal combination of endmember classes within each pixel, MEMM$_s$ incorporates the sparsity constraint to select the optimal combination. This prevents a larger number of endmember classes to be selected for each pixel.

\subsubsection{MEMM and pixel-wise endmember variability models}
By denoting $\mathbf{\tilde{M}}_i=\mathbf{E}\mathbf{B}_i$ the equivalent endmember matrix associated with the $i$th pixel, MEMM models the observed pixel spectra as
\begin{eqnarray}
	\mathbf{y}_i=\mathbf{\tilde{M}}_i\mathbf{a}_i+\mathbf{n}_i
\end{eqnarray}
where $\mathbf{\tilde{M}}_i$ can be interpreted as a set of $K$ spatially varying endmember spectra. This approach has been also adpoted in recent works to incorporate endmember variability as additive factors \cite{Thouv2016}, multiplicative factors \cite{Drume2016,Imbir2017} or a combination of additive and multiplicative factors \cite{Hong2017}. In particular, when $N_1=\ldots,N_k=1$ in \eqref{eq:bundles}, the endmember bundles $\mathbf{E}_1,\ldots, \mathbf{E}_K$ are reduced to unique endmember spectra characterizing each class. The associated bundling coefficient matrix $\mathbf{B} = \mathrm{diag}[b_1,\ldots,b_K]$ is diagonal where each coefficient $b_k$ scales the corresponding endmember spectrum $\mathbf{E}_k$ ($k=1,\ldots,K$). Thus, MEMM generalizes the recently introduced  extended linear mixing model \cite{Drume2016}.

However, MEMM is different from the aforementioned methods since it resorts to \textit{a priori} information (i.e., endmember bundles) to model the endmember variability. More precisely, MEMM describes the admissible variability within an endmember class as the convex cone spanned by the corresponding bundles. As a consequence, \emph{per se}, MEMM offers an adaptive description of the spectral variability even when pre-defined endmember bundles do not completely capture this variability within each class.

\subsubsection{MEMM and sparsity-based unmixing methods}
By setting $\mathbf{r}_i=\mathbf{B}_i\mathbf{a}_i$, MEMM in \eqref{eq:MEMM_model} can be rewritten as
\begin{eqnarray}
	\mathbf{y}_i=\mathbf{E}\mathbf{r}_i+\mathbf{n}_i
\end{eqnarray}
similarly to the existing models discussed in Section \ref{section2}. The main difference is that MEMM enables the multiple abundances $\mathbf{r}_i$ to be decomposed into bundling coefficients $\mathbf{B}_i$ within each class and abundances $\mathbf{a}_i$, resulting into a bi-layer description of the abundances. This hierarchical decomposition has been also adopted by unmixing methods based on multilayer nonnegative matrix factorization (MLNMF) \cite{Rajab2015}. However, each layer induced by MEMM (i.e., bundling matrix and abundance vector) has a clear and meaningful role. Moreover, when the existing methods impose ASC onto $\mathbf{r}_i$, they assume that mixed spectra belong to the simplex spanned by the endmember bundles. However, this is a limited assumption since observed spectra may be outside the simplex, e.g., when affected by variations in illumination. Conversely, MEMM imposes ASC only onto the abundances $\mathbf{a}_i$ of endmember classes and enables the bundling coefficients $\mathbf{B}_i$ to scale the endmember signature, e.g.,  to capture variability induced by varying illumination (see experiments in Section \ref{section6}). In addition,
MEMM complements this bi-layer hierarchy with a twofold structured, physically-motivated sparsity imposed on the multiple abundance vector $\mathbf{r}_i$. This bilevel sparsity has the significant advantage of reducing overfitting and one may expect a significant improvement of stability and interpretability of the abundance estimates.

Finally, the intrinsic structure of MEMM is also similar to the recently developed methods based on robust constrained matrix factorization \cite{Akhta2017} and kernel archetypoid analysis \cite{Zhao2017}. These methods model a set $\mathbf{Y} = \left[\mathbf{y}_1,\ldots,\mathbf{y}_P\right] \in \mathbb{R}^{L \times P}$ of $P$ pixel spectra as
\begin{eqnarray}
	\mathbf{Y}=\mathbf{Y}\mathbf{C}\mathbf{A}+\mathbf{N}
\end{eqnarray}
where $\mathbf{C} \in \mathbb{R}^{P \times K}$ is a matrix gathering a set of coefficients, $\mathbf{A} = \left[\mathbf{a}_1,\ldots,\mathbf{a}_P\right] \in \mathbb{R}^{K \times P}$ is the abundance matrix and $\mathbf{N} = \left[\mathbf{n}_1,\ldots,\mathbf{n}_P\right] \in \mathbb{R}^{L \times N}$ is the error and noise matrix. Sparsity (induced by ASC or the use of $\ell_0$-pseudonorm) and nonnegativity constraints are imposed onto each column of $\mathbf{C}$ and $\mathbf{Y}\mathbf{C}$ can be interpreted as synthetic endmember spectra. These methods use the subset of whole image pixels to generate synthetic endmember spectra that are fixed within each image. On the other hand, MEMM uses the subset of endmember bundles to generate synthetic endmember spectra that may be different for each pixel.

\section{MEMM-based unmixing algorithm}\label{sec:algo}
Unmixing according to the proposed MEMM can be formulated as the minimization problem
\begin{eqnarray}\label{eq_sep}
\begin{aligned}
	& \min\limits_{\mathbf{B}_i,\mathbf{a}_i}\frac{1}{2}\Vert\mathbf{E}\mathbf{B}_i\mathbf{a}_i-\mathbf{y}_i\Vert^2_2 \\
    \text{s.t.}\   \forall k,\forall i, & \ a_{ki} \geq 0,\quad  \sum_{k=1}^{K}a_{ki}=1,\quad \Vert\mathbf{a}_i\Vert_0 \leq v, \\
    &  \mathbf{B}_i \succeq 0,\quad \Vert\mathbf{B}_i\Vert_0 \leq s.
\end{aligned}
\end{eqnarray}
This minimization problem is similar to the double sparsity-inducing method proposed in \cite{Rubin2010}. Using an alternative formulation, the minimization problem can be written as the following non-convex minimization problem:
\begin{equation}
	\min\limits_{\mathbf{B}_i,\mathbf{a}_i} \mathcal{J}\left(\mathbf{B}_i,\mathbf{a}_i\right)=\left\lbrace f(\mathbf{B}_i,\mathbf{a}_i) + h(\mathbf{B}_i)+g(\mathbf{a}_i) \right\rbrace
\end{equation}
with
\begin{eqnarray}
     f(\mathbf{B}_i,\mathbf{a}_i)&=\frac{1}{2}\Vert\mathbf{E}\mathbf{B}_i\mathbf{a}_i-\mathbf{y}_i\Vert^2_2\\
     h(\mathbf{B}_i)&=\iota_{\mathbb{R}_+}(\mathbf{B}_i) + \lambda_b\Vert\mathbf{B}_i\Vert_0\\
     g(\mathbf{a}_i)&=\iota_{\mathbb{S}}(\mathbf{a}_i) + \lambda_a\Vert\mathbf{a}_i\Vert_0
\end{eqnarray}
where $\lambda_a$ and $\lambda_b$ are parameters which control the balance between the data fitting term and the sparse regularizations, $\iota_{\mathcal{C}}(\mathbf{x})$ is the indicator function on the set $C$ (i.e., $\iota_{\mathcal{C}}(\mathbf{x})=0$ when $\mathbf{x} \in \mathcal{C}$ whereas $\iota_{\mathcal{C}}(\mathbf{x})=\infty$ when $\mathbf{x} \notin \mathcal{C}$), and $\mathbb{S}$ is the simplex defined by the ASC and ANC. Solving this optimization problem is challenging since the  regularization functions $h$ and $g$ are nonconvex and nonsmooth. However, it can be tackled thanks to the proximal alternating linearized minimization (PALM)  \cite{Bolte2014}. With guarantees to converge to a critical point, PALM iteratively updates the parameters $\mathbf{a}_i$ and $\mathbf{B}_i$ by alternatively minimizing the objective function with respect to (w.r.t.) these parameters, i.e., by solving the following proximal problems
\begin{eqnarray}
\begin{aligned}
	\mathbf{B}_i^{(t+1)} \in & \min\limits_{\mathbf{B}_i} \left\lbrace h(\mathbf{B}_i) + \langle \mathbf{B}_i-\mathbf{B}_i^{(t)},\nabla_{\mathbf{B}_i}f(\mathbf{B}_i^{(t)},\mathbf{a}_i^{(t)})  \rangle \right.\\
    &\left. + \frac{c_t}{2}\Vert\mathbf{B}_i-\mathbf{B}_i^{(t)}\Vert_2^2 \right\rbrace\\
    \mathbf{a}_i^{(t+1)} \in & \min\limits_{\mathbf{a}_i} \bigg\{ g(\mathbf{a}_i) + \langle \mathbf{a}_i-\mathbf{a}_i^{(t)},\nabla_{\mathbf{a}_i}f(\mathbf{B}_i^{(t+1)},\mathbf{a}_i^{(t)}) \rangle \}. \\
    &\left. + \frac{d_t}{2}\Vert\mathbf{a}_i-\mathbf{a}_i^{(t)}\Vert_2^2 \right\rbrace
\end{aligned}
\end{eqnarray}
The pseudocode for MEMM is shown in Algorithm~\ref{algorithm_memm} and these two steps are described in what follows.

\subsection{Optimization w.r.t. $\mathbf{B}_i$}
To optimize only w.r.t. the diagonal entries in $\mathbf{B}_i$, the objective function can be rewritten with the following decomposition
\begin{eqnarray}
\begin{aligned}
	f(\mathbf{b}_i,\mathbf{a}_i)& =\frac{1}{2}\Vert\mathbf{U}_i\mathbf{b}_i-\mathbf{y}_i\Vert^2_2\\
    h(\mathbf{b}_i)& =\iota_{\mathbb{R}_+}(\mathbf{b}_i) + \lambda_b\Vert\mathbf{b}_i\Vert_0
\end{aligned}
\end{eqnarray}
where
\begin{eqnarray*}
\begin{aligned}
	\mathbf{U}_i&=\left[\mathbf{E}_1 \odot a_{1i}| \cdots| \mathbf{E}_K\odot a_{Ki}\right]\\
    \mathbf{b}_i &= \left[ \mathbf{b}_{1i}^T, \mathbf{b}_{2i}^T, \cdots, \mathbf{b}_{Ki}^T\right]^T.
\end{aligned}
\end{eqnarray*}
This leads to the following updating rule
\begin{eqnarray*}
	\min\limits_{\mathbf{b}_i} \left\lbrace h(\mathbf{b}_i) + \frac{c_{t}}{2}\Vert\mathbf{b}_i-(\mathbf{b}_i^{(t)}-\frac{1}{c_{t}}\nabla_{\mathbf{b}}f(\mathbf{b}_i^{(t)},\mathbf{a}_i^{(t)}))\Vert_2^2 \right\rbrace
\end{eqnarray*}
where $\nabla_{\mathbf{b}_i}f(\mathbf{b}_i^{(t)},\mathbf{a}_i^{(t)})=\mathbf{U}_i^T\left(\mathbf{U}_i \mathbf{b}_i - \mathbf{y}_i\right)$.
Using similar computations as in \cite{Bolte2014}, this can be conducted as
\begin{eqnarray}
	\begin{aligned}
	\mathbf{b}_i^{(t+1)} & \in \text{prox}_{c_t/\lambda_b}^h(\mathbf{b}_i^{(t)}-\frac{1}{c_t}\nabla_{\mathbf{b}_i}f(\mathbf{b}_i^{(t)},\mathbf{a}_i^{(t)}))\\
    \end{aligned}
\end{eqnarray}
where $c_{t}=\gamma_m\Vert\mathbf{U}_i^T\mathbf{U}_i\Vert_{\mathrm{F}}$ represents a step size for each iteration. The proximal operator associated with $f$ can be computed using the approach \cite{Bolte2014}. Finally, the bundling matrix $\mathbf{B}_i$ can be reconstructed as $\mathbf{B}_i=\text{blkdiag}(\mathbf{b}_i)$ where $\text{blkdiag}(\cdot)$ generates the block diagonal matrix $\mathbf{B}_i$ from the vector $\mathbf{b}_i$.\\

\subsection{Optimization with respect to $\mathbf{a}_i$}
To optimize w.r.t. $\mathbf{a}_i$, the objective function can be rewritten using the decomposition
\begin{eqnarray}
	 f(\mathbf{B}_i,\mathbf{a}_i)=&\frac{1}{2}\Vert\mathbf{\tilde{M}}_i\mathbf{a}_i-\mathbf{y}_i\Vert^2_2\\
    g(\mathbf{a}_i)=&\iota_{\mathbb{S}}(\mathbf{a}_i) + \lambda_a\Vert\mathbf{a}_i\Vert_0
\end{eqnarray}
where $\mathbf{\tilde{M}}_i=\mathbf{E}\mathbf{B}_i$. Thus, updating the abundance vector can be formulated as
\begin{eqnarray*}
	\min\limits_{\mathbf{a}_i} \left\lbrace g(\mathbf{a}_i) + \frac{d_{t}}{2}\left\|\mathbf{a}_i-\left(\mathbf{a}_i^{(t)}-\frac{1}{d_{t}}\nabla_{\mathbf{a}_i}f(\mathbf{B}_i^{(t+1)},\mathbf{a}_i^{(t)})\right)\right\|^2 \right\rbrace
\end{eqnarray*}
where $\nabla_{\mathbf{a}_i}f(\mathbf{B}_i^{(t+1)},\mathbf{a}_i^{(t)})=\mathbf{\tilde{M}}_i^T\left(\mathbf{\tilde{M}}_i \mathbf{a}_i - \mathbf{y}_i\right)$. Using the proximal operator, this can be written as
\begin{eqnarray}
	\begin{aligned}
	\mathbf{a}_i^{(t+1)} & \in \text{prox}_{d_{t}/\lambda_a}^g\left(\mathbf{a}_i^{(t)}-\frac{1}{d_{t}}\nabla_{\mathbf{a}_i}f(\mathbf{B}_i^{(t+1)},\mathbf{a}_i^{(t)})\right)
    \end{aligned}
\end{eqnarray}
where $d_{t}=\gamma_a\Vert\mathbf{\tilde{M}}_i^T\mathbf{\tilde{M}}_i\Vert_{\mathrm{F}}$ represents a step size for each iteration. Moreover the proximal mapping associated with $g$  can be  performed using the method developed in \cite{Anast2013}.

\begin{algorithm}
\caption{Algorithm for MEMM-based unmixing}\label{algorithm_memm}
\begin{algorithmic}[1]
\State $\mathbf{Input}: \mathbf{y}_i,\mathbf{E}$
\State \textbf{Initialization}: $\mathbf{a}_i^{(0)}$ and $\mathbf{B}_i^{(0)}$.
\State Set $\mathbf{r}_i^{(0)}$ using an unmixing method (e.g. FCLS).
\State $\mathbf{a}_i^{(0)}=\mathbf{G}^T\mathbf{r}_i^{(0)}$
\State $\forall_k, \mathbf{b}_{ik}^{(0)}=(\mathbf{g}_k\odot\mathbf{r}_i^{(0)}) \oslash a_{ki}^{(0)}$
\State \textbf{Main procedure}:
\While{ the stopping criterion is not satisfied}
\State $\mathbf{b}_i^{(t+1)} \leftarrow  \text{prox}_{c_t/\lambda_b}^h(\mathbf{b}_i^{(t)}-\frac{1}{c_{t}}\nabla_{\mathbf{b}_i}f(\mathbf{b}_i^{(t)},\mathbf{a}_i^{(t)}))$
\State $\mathbf{B}_i^{(t+1)}=\text{blkdiag}(\mathbf{b}_i^{(t+1)})$
\State $\mathbf{a}_i^{(t+1)} \leftarrow  \text{prox}_{d_{t}/\lambda_a}^g(\mathbf{a}_i^{(t)}-\frac{1}{d_{t}}\nabla_{\mathbf{a}_i}f(\mathbf{B}_i^{(t+1)},\mathbf{a}_i^{(t)}))$
\EndWhile
\State $\mathbf{Output}: \mathbf{a}_i^{(t+1)},\mathbf{B}_i^{(t+1)}$
\end{algorithmic}
\end{algorithm}

\subsection{Initialization and stopping rule}
MEMM requires initial estimates $\mathbf{a}_i^{0}$ and $\mathbf{B}_i^{0}$ of the abundance vector and bundling matrix, respectively. To do so, first, MEMM estimates a multiple abundance vector $\mathbf{r}_i^{(0)}$ using a state-of-the-art LMM-based unmixing method  (e.g., FCLS, see line 3). Then an initial estimate of the single abundance vector $\mathbf{a}_i^{(0)}$ is computed according to \eqref{summed} (see line 4). Finally, the bundling matrix $\mathbf{B}_i^{(0)}$ is arbitrarily initialized as the corresponding scaling factor (see line 5, where $\oslash$ stands for the element-wise division). Once initial estimates have been obtained, $\mathbf{a}_i^{(t+1)}$ and $\mathbf{b}^{(t+1)}$ are iteratively updated in lines 8--10. The algorithm stops when the difference between updated and previous values of the objective function $f(\mathbf{B}_i,\mathbf{a}_i)$ is smaller than a predetermined threshold.

\begin{figure*}
        \centering
        \includegraphics[width=\linewidth]{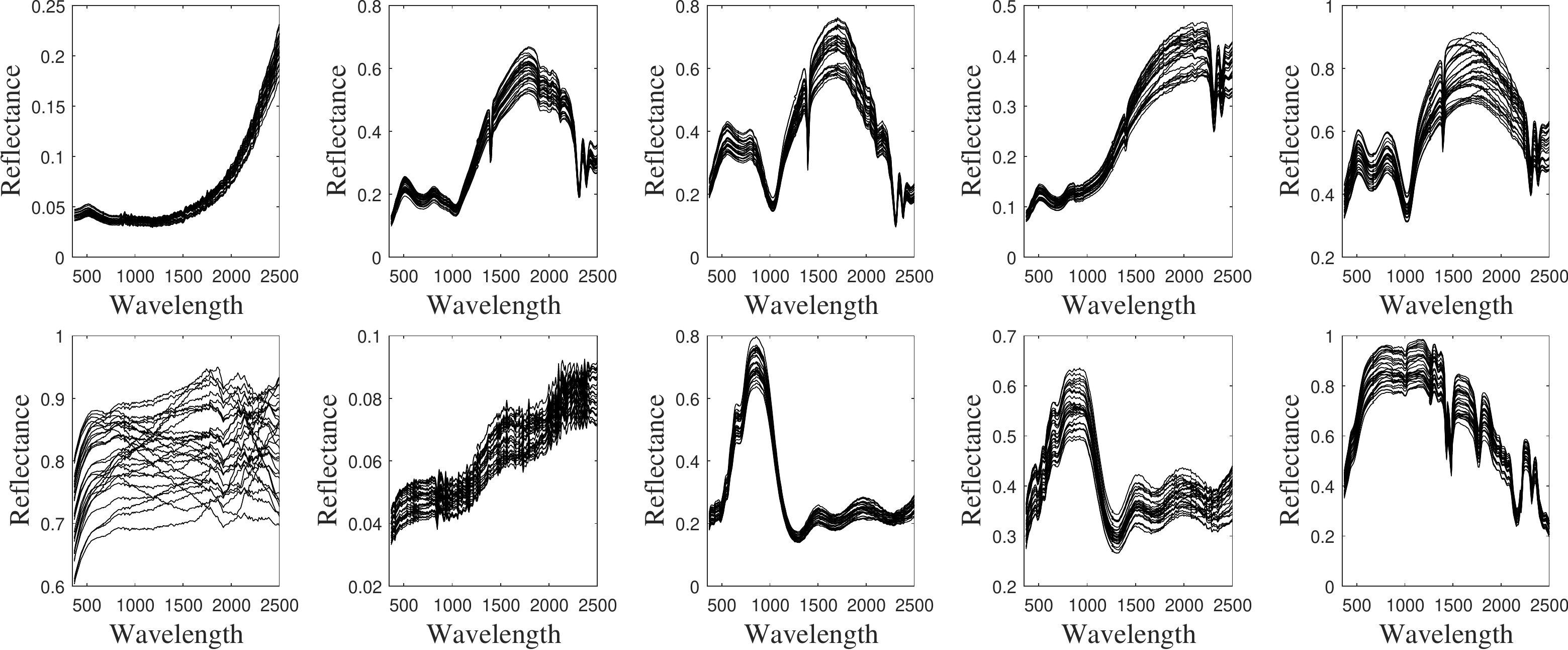}
        \caption{Synthetically generated endmember bundles.}\label{fig:endm_b_sim}
\end{figure*}

\section{Experiments using simulated data}\label{section4}
First, the relevance of the proposed MEMM and its variant MEMM$_{s}$ has been evaluated thanks to experiments conducted on simulated datasets.

\subsection{Generation of bundles}
First, $K=10$ spectra were selected from the USGS spectral library. The $10$ spectra were chosen so that the minimum angle between any two spectra was larger than $5^\circ$. This pruning prevented synthetic endmember bundles to overlap each other. Second, endmember bundles $\mathbf{E}_k$ ($k=1,\ldots,K$)  were designed by randomly generated $N_k=30$ endmember spectra for each endmember bundle using the approach proposed in \cite{Thouv2016a}. These bundles, depicted in Fig. \ref{fig:endm_b_sim},  were used for generating the following two simulated data.

\subsubsection{Simulated dataset 1 (SIM1)}
The first dataset, referred to as SIM1 in what follows, was generated using MESMA. A mixed spectrum was generated using the following 5 steps. First, the number of endmember classes $K$ was randomly determined in the set ($\{1,\ldots,5\}$). Second, a random combination of endmember classes was selected. Third, one spectrum within each selected endmember class was randomly chosen. Forth, the abundances of the selected endmember classes were randomly generated using a Dirichlet distribution to jointly ensure ANC and ASC. Finally, a mixed spectrum was generated by a linear combination of endmember spectra of the selected endmember classes and the randomly generated abundances. A set of $P=100$ mixed spectra were generated in this study. Different amounts of additive Gaussian noise with corresponding signal-to-noise ratios of $50$dB, $40$dB and $30$dB were considered to the mixed spectra.

\subsubsection{Simulated dataset 2 (SIM2)}
SIM2 was generated using MEMM. A mixed spectrum was generated similarly as done in SIM1. The main difference is the bundling coefficients in MEMM. In order to generate the bundling coefficients, the number of spectra $N_k$ was randomly chosen in the set ($\{1,\ldots,5\}$). The bundling coefficients of the randomly selected spectra were generated from a Dirichlet distribution. A mixed spectrum was generated by a linear combination of spectra, the bundling coefficients and the abundances. As for SIM1, $P=100$ mixed spectra were generated and different amounts of Gaussian noise were also added to SIM2.

\subsection{Compared methods}
MEMM and MEMM$_s$ were compared with other 5 methods that incorporate endmember bundles and promote sparsity: fully constrained least squares (FCLS) \cite{Biouc2010}, sparse unmixing by variable splitting and augmented Lagrangian (SUnSAL) \cite{Iorda2011}, alternating angle minimization (AAM) \cite{Heyle2016a} and methods based on group lasso and elitist lasso \cite{Meyer2016}. Note that FCLS and SUnSAL can incorporate endmember bundles by considering the model in (\ref{eq:4}). FCLS and SUnSAL were chosen for comparing the proposed methods with most widely used unmixing methods that promote sparsity. AAM was selected for comparison because it is a latest variant of MESMA and it is more computationally efficient than MESMA while achieving good performance. Group lasso and elitist lasso were also included for comparison because they can incorporate the structure of endmember bundles as well as MEMM. MEMM, MEMM$_s$ and the methods based on group lasso and elitist lasso require initial estimates of abundances. In order to fairly compare methods, this study used abundances estimated by FCLS for the initial estimates required for the methods. SUnSAL, the methods based on group lasso or elitist lasso and MEMM$_s$ require a parameter $\lambda_r$ or $\lambda_a$ controlling sparsity regularization. MEMM requires two parameters $\lambda_b$ and $\lambda_a$. In order to fairly compare the methods, these parameters were empirically determined in the set $(0.0001, 0.001, 0.01, 0.1, 1, 5)$ for each simulated data so that the selected values produced the highest SRE$_a$. Parameter sensitivity of the proposed method is reported in the supplementary document \cite{Uezat2018sup}. Finally, computational time were also discussed.

\subsection{Performance criteria}
The main objective of this experiment was to assess the algorithm performance while selecting a combination of endmember classes and spectra, and estimating abundances corresponding to endmember classes and spectra. Three criteria were chosen for quantitative validation of the methods. To evaluate the quality of the reconstruction, one defines the signal-to-reconstruction error (SRE) per endmember class and per endmember spectrum as \cite{Iorda2014a}
\begin{equation}
\begin{aligned}
\text{SRE}_a \equiv \mathbb{E}\left[\Vert \mathbf{a} \Vert_2^2 \right]/\mathbb{E}\left[\Vert \mathbf{a}-\hat{\mathbf{a}} \Vert_2^2 \right]\\
\text{SRE}_r \equiv \mathbb{E}\left[\Vert \mathbf{r} \Vert_2^2 \right]/\mathbb{E}\left[\Vert \mathbf{r}-\hat{\mathbf{r}} \Vert_2^2 \right]\\
\end{aligned}
\end{equation}
where $\mathbf{a}$ and $\hat{\mathbf{a}}$ are the actual and estimated abundance vectors of all pixels, $\mathbf{r}$ and $\hat{\mathbf{r}}$ are the actual and estimated multiple abundance vectors of all pixels.

Second, the number of nonzero abundances were used to evaluate the sparsity level per endmember class and per endmember spectrum recovered by the methods, i.e.,  \cite{Iorda2014a,Chen2016}
\begin{equation}
\begin{aligned}
\text{SL}_a \equiv \frac{1}{P}\sum_{i=1}^{P} \Vert \hat{\mathbf{a}}_i \Vert_0\\
\text{SL}_r \equiv \frac{1}{P}\sum_{i=1}^{P} \Vert \hat{\mathbf{r}}_i \Vert_0.
\end{aligned}
\end{equation}
As in \cite{Iorda2014a}, abundances smaller than $10^{-4}$ were considered as \textit{zero abundances}.

Finally, to validate the performance in selecting a relevant combination of endmember classes or spectra, one defines the distance between the two actual and estimated supports (DIST) \cite{Elad2010,Chen2016}
\begin{equation}
\begin{aligned}
\text{DIST}_a \equiv \frac{1}{P}\sum_{i=1}^{P} \frac{\max\left(\vert \mathcal{S}_{i}^a \vert, \vert \hat{\mathcal{S}}_{i}^a \vert\right) - \vert \mathcal{S}_{i}^a \cap \hat{\mathcal{S}}_{i}^a\vert}{\max\left(\vert \mathcal{S}_{i}^a \vert, \vert \hat{\mathcal{S}}_{i}^a \vert\right)}\\
\text{DIST}_r \equiv \frac{1}{P}\sum_{i=1}^{P} \frac{\max\left(\vert \mathcal{S}_{i}^r \vert, \vert \hat{\mathcal{S}}_{i}^r \vert\right) - \vert \mathcal{S}_{i}^r \cap \hat{\mathcal{S}}_{i}^r\vert}{\max\left(\vert \mathcal{S}_{i}^r \vert, \vert \hat{\mathcal{S}}_{i}^r \vert\right)}\\
\end{aligned}
\end{equation}
where $\mathcal{S}$ and $\hat{\mathcal{S}}$ are true and estimated support sets (i.e., indexes of nonzero values), $\vert \mathcal{S} \vert$ represents the total number of elements in the set $\mathcal{S}$ and $\cap$ stands for the intersection operator. The figures of merit DIST$_a$ and DIST$_r$ evaluated the distance between two supports of endmember classes and the distance between two supports of endmember spectra, respectively.

\begin{table}[h!]
    \caption{SRE per endmember class ($\text{SRE}_a$).}\label{table:sre_c}
\resizebox{\columnwidth}{!}{
\begin{tabular}{c||c||ccccccc}									
\toprule									
	&SNR	&FCLS	&AAM	&SUnSAL	&Group lasso	&Elitist lasso	&MEMM$_s$	&MEMM	\\
\midrule									
\multirow{3}{*}{SIM1}	&30dB	&18.0507	&18.7656	&18.9285	&\textbf{19.4215}	&19.2333	 &14.5739	&18.6904	\\
	&40dB	&22.9074	&\textbf{25.8112}	&23.0496	&23.589	&23.303	&14.8229	&23.3532	\\
	&50dB	&27.358	&\textbf{33.8279}	&27.5013	&27.4184	&26.8979	&14.1579	&28.0017	 \\
\midrule									
\multirow{3}{*}{SIM2}	&30dB	&16.1865	&15.9501	&16.5182	&\textbf{18.0736}	&16.4315	 &12.1614	&16.4785	\\
	&40dB	&21.6942	&19.4906	&21.6671	&22.0263	&21.3809	&13.6682	 &\textbf{22.1381}	\\
	&50dB	&25.6617	&22.9952	&25.5168	&25.9215	&24.7702	&14.9405	 &\textbf{26.5442}	\\
\bottomrule									
\end{tabular}																
}
 \end{table}

\begin{table}[h!]
    \caption{SRE per endmember spectrum ($\text{SRE}_r$).}\label{table:sre_s}
\resizebox{\columnwidth}{!}{
\begin{tabular}{c||c||ccccccc}									
\toprule									
	&SNR	&FCLS	&AAM	&SUnSAL	&Group lasso	&Elitist lasso	&MEMM$_s$	&MEMM	\\
\midrule									
\multirow{3}{*}{SIM1}	&30dB	&1.1752	&\textbf{2.2035}	&1.7234	&2.0398	&1.5188	&-0.4570	 &0.7692	\\
	&40dB	&2.8611	&\textbf{4.6569}	&3.0262	&3.0357	&2.9957	&0.5171	&2.7653	\\
	&50dB	&3.6312	&\textbf{12.7796}	&3.7692	&3.7591	&3.7576	&0.1700	&3.5721	\\
\midrule									
\multirow{3}{*}{SIM2}	&30dB	&1.0084	&-0.2324	&1.4405	&\textbf{1.8935}	&1.3043	&-2.0312	 &0.4313	\\
	&40dB	&2.3323	&1.4226	&2.4976	&\textbf{2.5741}	&2.48	&-2.1441	&2.2462	\\
	&50dB	&3.2175	&3.3104	&3.3273	&3.3266	&\textbf{3.3666}	&-3.0151	&3.2304	\\
\bottomrule									
\end{tabular}															
}
 \end{table}

\subsection{Results}
SRE per class was calculated for each method  and reported in Table \ref{table:sre_c}. For SIM1 with $50$dB, AAM performed best among all methods. The performance of AAM, however, was degraded as SNR became lower. For data with $30$dB, the results derived from AAM were  worse than those derived from SUnSAL, elitist lasso and group lasso. In addition, AAM performed poorly compared with other methods in SIM2. This showed that the MESMA-based approach (AAM) was less effective when given endmember bundles did not completely represent endmember variability present in the data and SNR of the data was low ($<40$dB). This finding was also observed in \cite{Uezat2016}. MEMM produced better results for SIM2 with $50$dB  than the other sparsity-methods and produced comparable results with $40$dB and $30$dB. MEMM$_s$ performed poorly, compared with all methods. SRE per spectrum was also calculated from each method  and reported in Table \ref{table:sre_s}. Compared with SRE per class, SRE per spectrum was very low for all methods. This showed that the exact recovery of multiple abundances $\mathbf{r}_i$ was challenging under conditions where a large number of endmember spectra were present within each class.

\begin{table}[h!]
    \caption{Sparsity level per endmember class ($\text{SL}_a$). Reference: $\text{SL}_a=2.01$ in SIM1 and $\text{SL}_a=2.27$ in SIM2.)}\label{table:nonneg_c}
\resizebox{\columnwidth}{!}{
\begin{tabular}{c||c||ccccccc}			
\toprule									
	&SNR	&FCLS	&AAM	&SUNSAL	&Group lasso	&Elitist lasso	&MEMM$_s$	&MEMM	\\
\midrule									
\multirow{3}{*}{SIM1}	&30dB	&4.8	&4.25	&3.32	&4.59	&6.83	&3.05	&\textbf{1.84}	\\
	&40dB	&4.62	&4.29	&3.82	&4.54	&5.64	&3.64	&\textbf{2.14}	\\
	&50dB	&4.34	&4.05	&4.24	&4.58	&5	&3.55	&\textbf{1.99}	\\
\midrule									
\multirow{3}{*}{SIM2}	&30dB	&5.02	&4.45	&3.56	&4.72	&6.63	&4.27	&\textbf{2.03}	\\
	&40dB	&5.21	&4.59	&4.16	&5.02	&5.6	&4.25	&\textbf{2.5}	\\
	&50dB	&4.97	&4.49	&4.87	&4.72	&5.54	&3.9	&\textbf{2.28}	\\
\bottomrule									
\end{tabular}														
}
 \end{table}

\begin{table}[h!]
    \caption{Sparsity level per endmember class ($\text{SL}_r$). Reference: $\text{SL}_r=2.01$ in SIM1 and $\text{SL}_r=2.39$ in SIM2.}\label{table:nonneg_s}
\resizebox{\columnwidth}{!}{
\begin{tabular}{c||c||ccccccc}								
\toprule									
	&SNR	&FCLS	&AAM	&SUNSAL	&Group lasso	&Elitist lasso	&MEMM$_s$	&MEMM	\\
\midrule		
\multirow{3}{*}{SIM1}	&30dB	&10.83	&4.25	&10.52	&43.12	&11.28	&\textbf{3.04}	&5.07	\\
	&40dB	&14.25	&4.29	&12.39	&29.88	&15.81	&\textbf{3.55}	&8.72	\\
	&50dB	&18.66	&4.05	&18.07	&23.57	&21.31	&\textbf{3.45}	&11.13	\\
\midrule									
\multirow{3}{*}{SIM2}	&30dB	&11.98	&4.45	&13.08	&66.64	&13.16	&\textbf{4.22}	&5.26	\\
	&40dB	&16.79	&4.59	&15.6	&45.99	&19.67	&\textbf{4.15}	&10.88	\\
	&50dB	&25.34	&4.49	&24.84	&65.97	&29.76	&\textbf{3.78}	&19.38	\\
\bottomrule									
\end{tabular}															
}
 \end{table}

The average number of nonzero abundances in $\mathbf{a}_i$ and $\mathbf{r}_i$ were shown in Table \ref{table:nonneg_c} and Table \ref{table:nonneg_s}, respectively. For the number of nonzero abundances per class, MEMM performed best among all methods for both SIM1 and SIM2. This showed that the sparse constraint used in MEMM successfully led to the selection of smaller numbers of endmember classes. For the number of nonzero abundances per spectrum, AAM and MEMM$_s$ produced the smaller number of nonzero abundances than other methods. This was because these methods selected at most one spectrum within each class and enforced greater sparsity when selecting endmember spectra.

\begin{table}[h!]
    \caption{Distance between actual and estimated supports per endmember class ($\text{DIST}_a$).}\label{table:dist_c}
\resizebox{\columnwidth}{!}{
\begin{tabular}{c||c||ccccccc}									
\toprule									
	&SNR	&FCLS	&AAM	&SUNSAL	&Group lasso	&Elitist lasso	&MEMM$_s$	&MEMM	\\
\midrule									
\multirow{3}{*}{SIM1}	&30dB	&0.5893	&0.5489	&0.3939	&0.5866	&0.6967	&0.3491	&\textbf{0.1195}	 \\
	&40dB	&0.5689	&0.5163	&0.4616	&0.5594	&0.6368	&0.4498	&\textbf{0.1265}	\\
	&50dB	&0.5137	&0.4828	&0.5035	&0.5479	&0.5795	&0.4180	&\textbf{0.0758}	\\
\midrule									
\multirow{3}{*}{SIM2}	&30dB	&0.5725	&0.5156	&0.3775	&0.5487	&0.6538	&0.5025	&\textbf{0.1843}	 \\
	&40dB	&0.5789	&0.4843	&0.4521	&0.5534	&0.5952	&0.4938	&\textbf{0.162}	\\
	&50dB	&0.5345	&0.4696	&0.5247	&0.5185	&0.5825	&0.4179	&\textbf{0.1033}	\\
\bottomrule									
\end{tabular}							
}
 \end{table}

\begin{table}[h!]
    \caption{Distance between actual and estimated supports per endmember spectrum ($\text{DIST}_r$).}\label{table:dist_s}
\resizebox{\columnwidth}{!}{
\begin{tabular}{c||c||ccccccc}									
\toprule									
	&SNR	&FCLS	&AAM	&SUNSAL	&Group lasso	&Elitist lasso	&MEMM$_s$	&MEMM	\\
\midrule									
\multirow{3}{*}{SIM1}	&30dB	&0.9167	&0.767	&0.853	&0.94	&0.8963	&\textbf{0.7493}	&0.8115	 \\
	&40dB	&0.8914	&\textbf{0.6707}	&0.8587	&0.9137	&0.8881	&0.7351	&0.774	\\
	&50dB	&0.8649	&\textbf{0.5622}	&0.8578	&0.8971	&0.8911	&0.6586	&0.7431	\\
\midrule									
\multirow{3}{*}{SIM2}	&30dB	&0.9039	&0.8426	&0.8476	&0.9332	&0.8915	&0.8729	&\textbf{0.8142}	 \\
	&40dB	&0.8791	&\textbf{0.7672}	&0.8525	&0.9167	&0.883	&0.8175	&0.7993	\\
	&50dB	&0.8738	&\textbf{0.6949}	&0.8696	&0.9172	&0.8816	&0.7277	&0.8095	\\
\bottomrule									
\end{tabular}																
}
 \end{table}

The errors in the support sets were shown in Table \ref{table:dist_c} and Table \ref{table:dist_s}. When estimating the support set per class, MEMM outperformed other methods for both SIM1 and SIM2. This showed that the double sparsity imposed by MEMM also led to the appropriate selection of the combination of endmember classes for each pixel. The performance of MEMM, however, was degraded when estimating the support set per spectrum. Other methods also performed poorly for the support set per spectrum. This shows that when multiple highly correlated endmember spectra are present within each class, the existing methods experience difficulty to select an optimal combination of endmember spectra.

\begin{table}[h!]
    \caption{Computational time for unmixing $P=100$ pixels using endmember bundles of $N=300$ endmember spectra.}\label{table:time}
\resizebox{\columnwidth}{!}{
\begin{tabular}{c|ccccccc}								
\toprule								
Data	&FCLS	&AAM	&SUNSAL	&Group lasso	&Elitist lasso	&MEMM$_s$	&MEMM	\\
\midrule								
SIM1	&0.3006	&517.7361	&\textbf{0.155}	&0.9517	&1.2694	&0.4846	&2.6625	\\
SIM2	&\textbf{0.1966}	&549.1244	&0.2616	&0.8426	&1.2122	&2.2311	&2.5705	\\
\bottomrule								
\end{tabular}								
}
 \end{table}

Finally, the computational times of all methods were shown in Table \ref{table:time}. MEMM was more computationally expensive than FCLS, SUnSAL, Group lasso and Elitist lasso. The proposed methods, however, were computationally cheaper than AAM because they did not need to test a large number of the combinations of endmember spectra.

\begin{figure}[h!]
        \begin{subfigure}[b]{0.5\columnwidth}
                \centering
                \includegraphics[width=.9\linewidth]{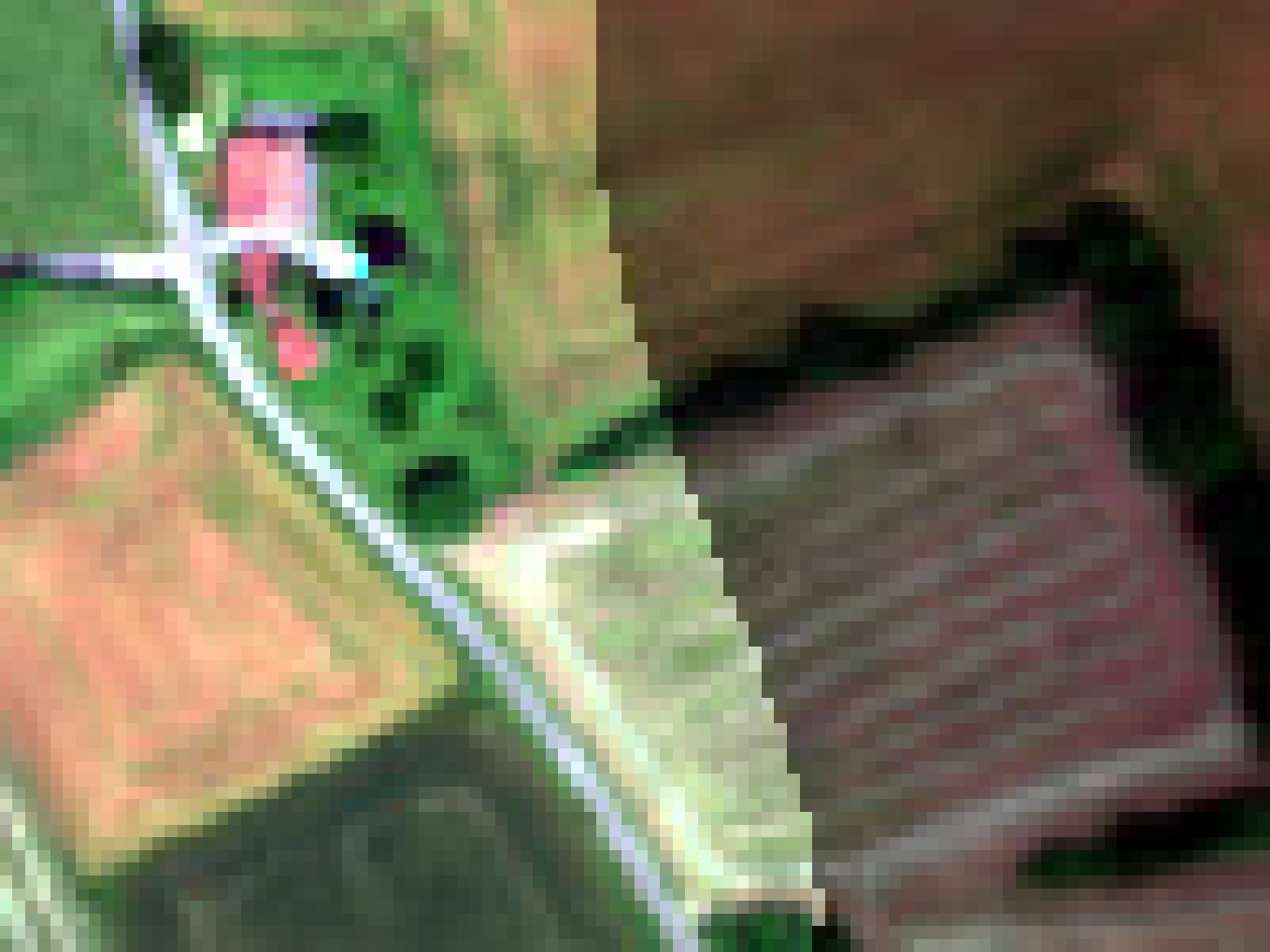}
                \caption{}
                \label{fig:real_rgb_1}
        \end{subfigure}%
        \begin{subfigure}[b]{0.5\columnwidth}
                \centering
                \includegraphics[width=.9\linewidth]{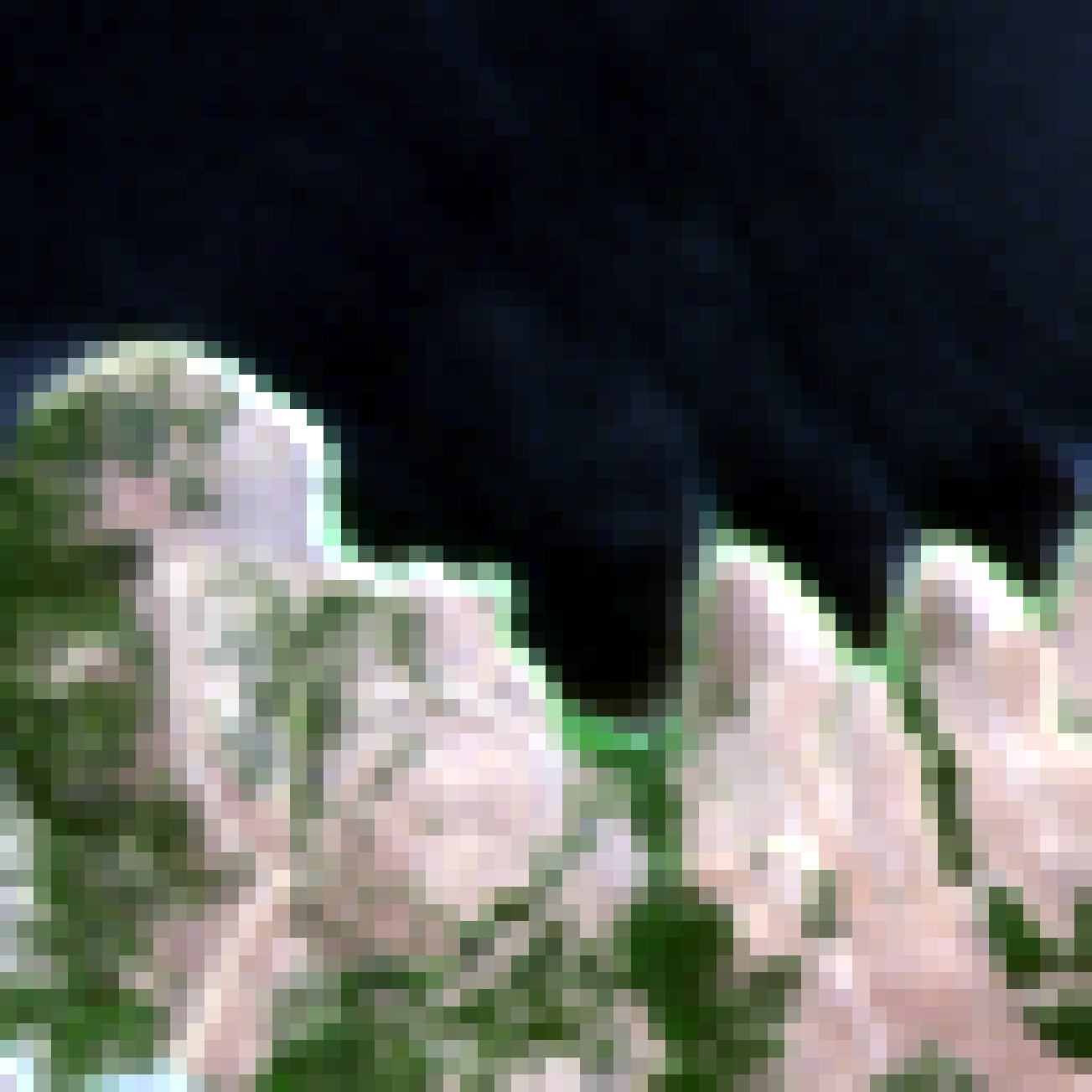}
                \caption{}
                \label{fig:real_rbg_2}
        \end{subfigure}
        \caption{Real hyperspectral images: (a) MUESLI image and (b) AVIRIS image.}\label{fig:rgb}
\end{figure}

\section{Experiments using real data}\label{section5}
\subsection{Description of real hyperspectral images}

\begin{figure*}
        \centering
        \includegraphics[width=\linewidth]{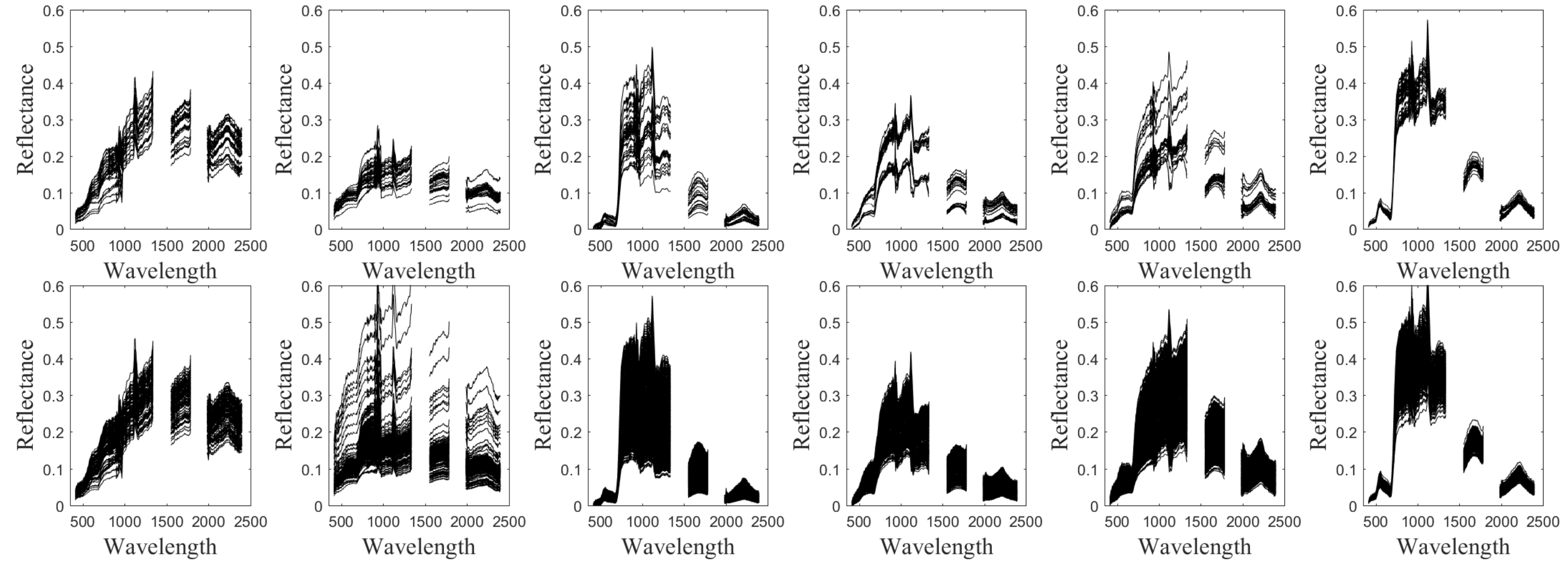}
        \caption{First row: endmember bundles used for unmixing the MUESLI image. Second row: endmember bundles generated by MEMM.}\label{fig:endm_b_real1}
\end{figure*}

The methods were finally compared on two real hyperspectral images. The first hyperspectral image was acquired in June 2016, over the city of Saint-Andr{\'e}, France, during the MUESLI airborne acquisition campaign. The image was composed of $415$ spectral bands. The spectral bands affected by noise (between $1.34-1.55\mu$m and $1.80-1.98\mu$m) were removed, leading to $L=345$ spectral bands.  In the image scene, spatially discrete objects were present. In this study, each spatially discrete region was assumed to be composed of a single endmember class. Endmember bundles were extracted from each region. As a consequence, large amounts of endmember variability were expected to be present within each spatially discrete region associated with a particular class and mixed pixels were expected to be located in the boundary of these spatially discrete regions. Moreover, the scene of interest was composed of two flight lines under significantly different illumination conditions, as shown in Fig. \ref{fig:real_rgb_1}. Thus, this image (referred to as MUESLI image) was used to evaluated whether the methods could accurately estimate abundances when large amounts of endmember variability were present. From this image, $K=6$ endmember bundles composed of a total of $N=180$ spectral signatures representing spatially discrete objects were extracted using the n-Dimensional Visualizer provided by the ENVI software. These bundles are represented in Fig. \ref{fig:endm_b_real1}(top). Some endmember classes present in the studied area were affected by different illumination conditions. Unlike simulated data, ground truth was not available. Estimated abundances were qualitatively validated by visual inspection of the abundance maps.

\begin{figure}
        \centering
        \includegraphics[width=.7\linewidth]{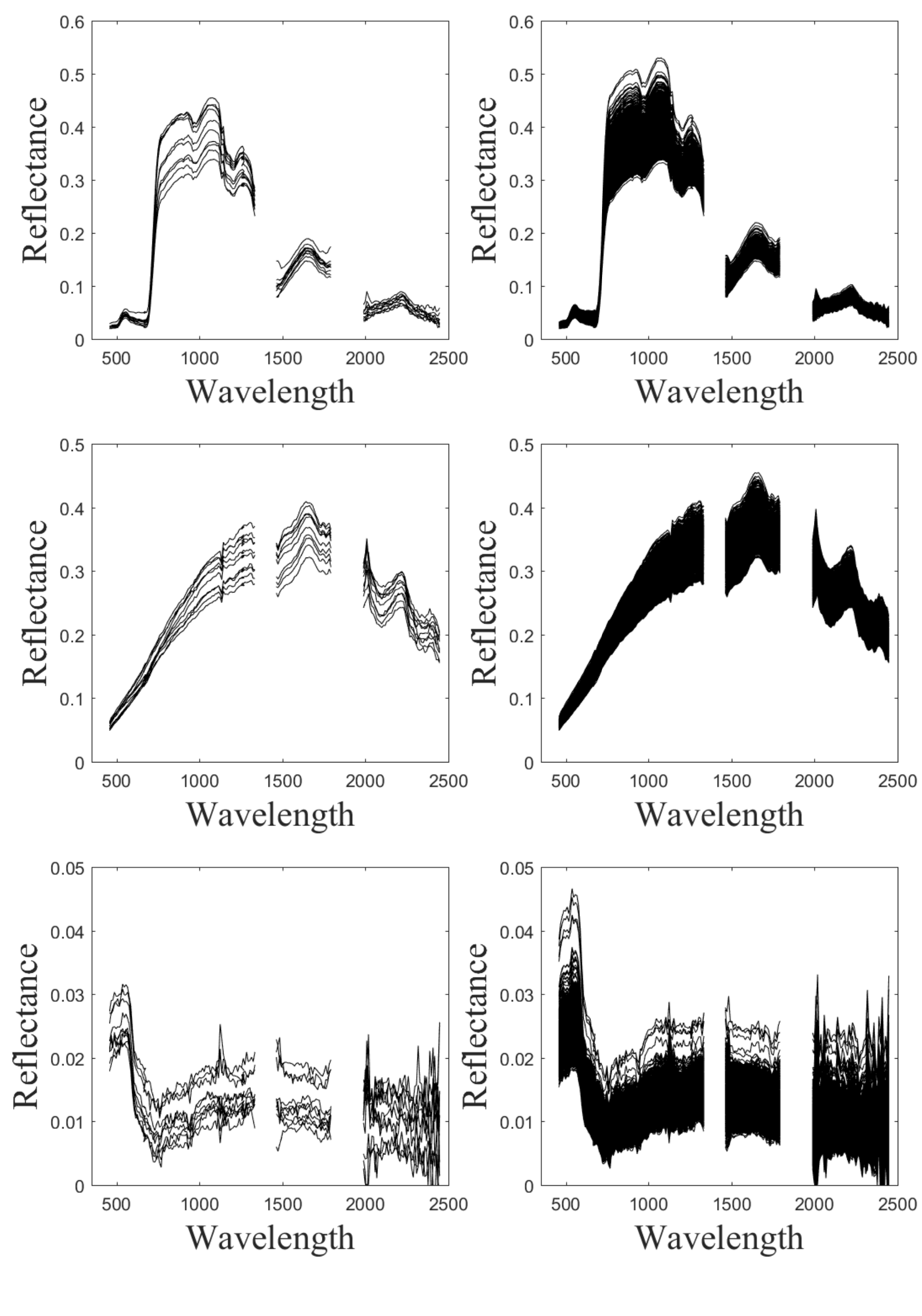}
        \caption{First column: endmember bundles used for unmixing the AVIRIS image. Second column: endmember bundles generated by MEMM.}\label{fig:endm_b_real2}
\end{figure}

The second image was acquired over Moffett Field, CA, USA, by the Airborne Visible/Infrared Imaging Spectrometer (AVIRIS). The image, depicted in Fig. \ref{fig:real_rbg_2}, initially comprised $224$ spectral bands. After the noisy spectral bands were removed, $L=178$ spectral bands remained. The area of interest, composed of a lake and a vegetated coastal area, was considered in many previous studies, e.g., to assess the performance of unmixing methods. Thus, this second image (referred to as AVIRIS image) was used to test whether the proposed method could perform at least as well as the existing methods to analyze this widely used image scene. As for the AVIRIS image, $K=3$ endmember bundles were extracted and qualitative validation was conducted because of the lack of ground truth. The extracted bundles are depicted in Fig. \ref{fig:endm_b_real2}(left).

\subsection{Results}
For both images, the proposed method was compared with FCLS, SUnSAL, AAM, the methods based on group lasso and elitist lasso. The parameters ($\lambda_r$, $\lambda_b$ and $\lambda_a$) required for the methods were empirically determined by qualitatively evaluating abundances derived from different values of the parameters. Finally, synthetic endmember bundles generated by MEMM were compared with endmember bundles initially used for unmixing.

\begin{figure*}
        \centering
        \includegraphics[width=\linewidth]{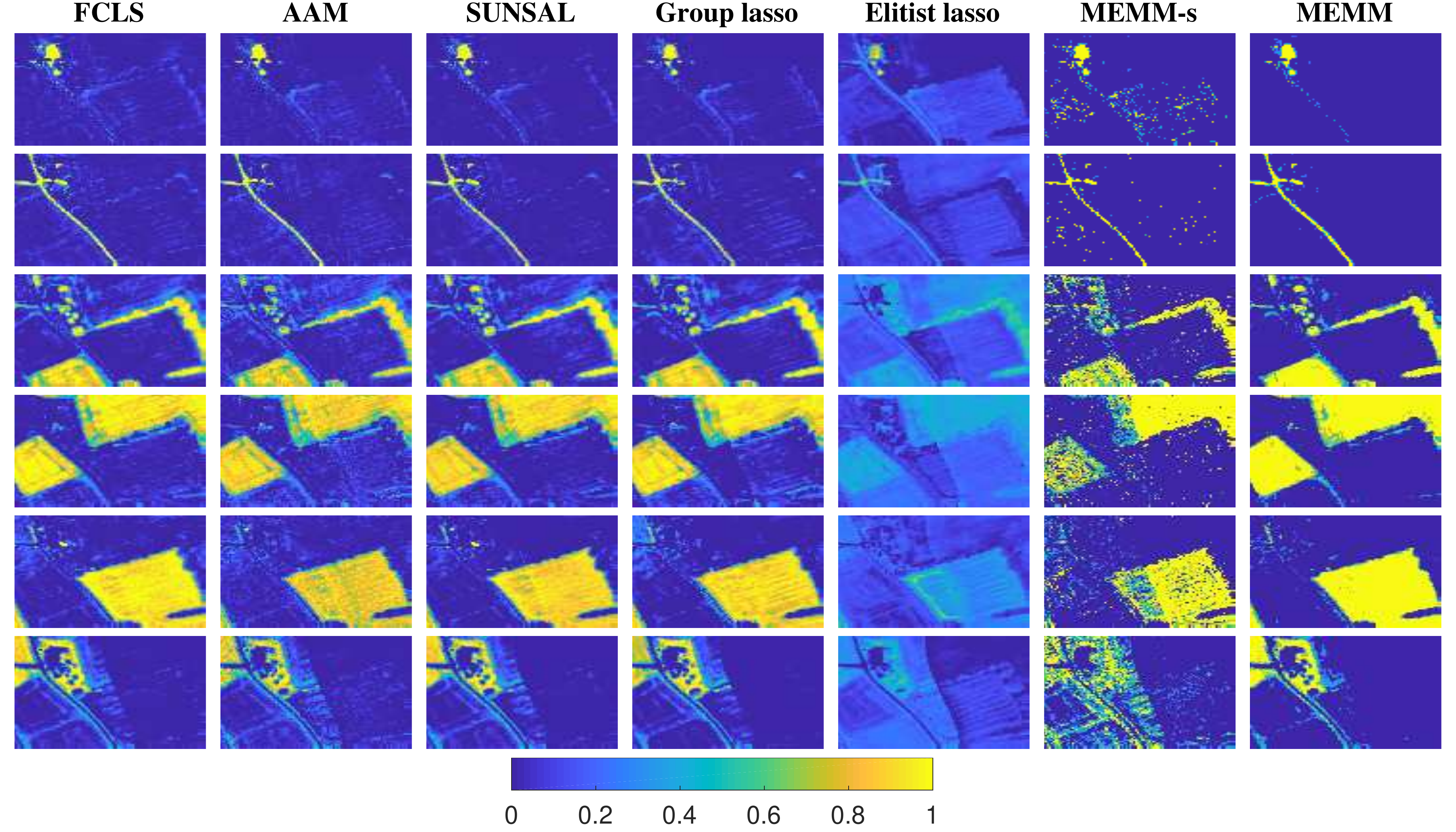}
        \caption{MUESLI image: estimated abundance maps. From top to bottom: building, road, shrub, crop land~1, crop land~2 and grass.}\label{fig:abun_all_real1}
\end{figure*}

Abundance maps estimated by the $7$ methods on the MUESLI image are depicted in Fig. \ref{fig:abun_all_real1}. These maps show that the abundances estimated by MEMM were more consistent at the boundary affected by the different illumination conditions. This showed that MEMM was more robust to the different illumination conditions than concurrent methods. The abundances estimated by MEMM were also high for each endmember class and showed less noisy. This suggested that MEMM also promoted more sparsity than other methods.

\begin{figure*}
        \centering
        \includegraphics[width=\linewidth]{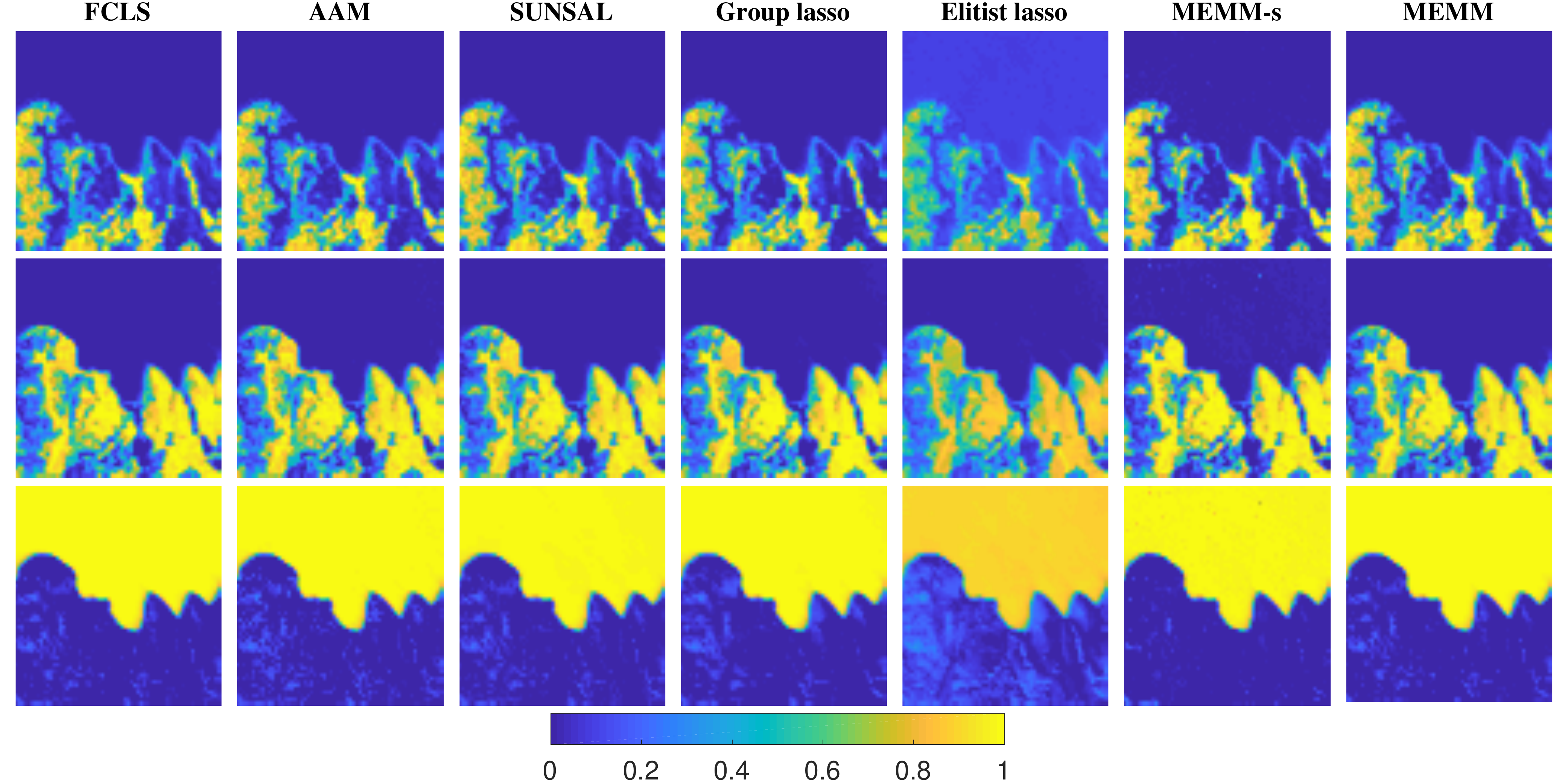}
        \caption{AVIRIS image: abundance maps. From top to bottom: vegetation, soil and water.}\label{fig:abun_all_real2}
\end{figure*}

Fig. \ref{fig:abun_all_real2} shows the abundance maps estimated for the AVIRIS image. All methods except elitist lasso generated similar abundances. Abundances estimated by elitist lasso were different because it was designed to use a larger number endmember classes for unmixing each pixel. MEMM produced similar abundances when compared with FCLS, group lasso. This showed that MEMM could perform at least as well as other sparsity-based methods to unmix this well-studied test site. MEMM$_s$, however, generated more noisy abundance maps. This showed that the initial estimates of abundances used in MEMM$_s$ did not lead to an optimal combination of endmember spectra and the optimal abundances.

Finally, the endmember bundles recovered by MEMM were compared with endmember bundles initially used to unmix both hyperspectral images, see Fig. \ref{fig:endm_b_real1} and Fig. \ref{fig:endm_b_real2}. In Fig. \ref{fig:endm_b_real1}, the synthetic endmember bundles filled the gaps that were present in the original endmember bundles. The extended endmember bundles showed more detailed spectral variability within each class in terms of both spectrum amplitudes and shapes also in Fig. \ref{fig:endm_b_real2}. This enabled MEMM to generate adaptive endmember spectra within each pixel and estimate more accurate abundances even when initial endmember bundles did not completely represent endmember variability.

\section{Conclusion}\label{section6}
This paper proposed a multiple endmember mixing model that bridges the gap between endmember bundle-based method and data driven-based methods. MEMM appeared to be superior to the existing methods as follows: \emph{i}) it incorporated endmember bundles to generate adaptive endmember spectra for each pixel, \emph{ii}) it had explicit physical meaning and generated hierarchical structure of endmember spectra, \emph{iii}) it imposed double sparsity for the selection of both endmember classes and endmember spectra. MEMM were tested and compared to the state-of-the-art methods using simulated data and real hyperspectral images. MEMM showed comparable results for estimating abundances while it outperformed other methods in terms of selecting a set of endmember classes within each pixel. This paper deeply focused on sparsity constraints for both bundling coefficients and abundances. However, other constraints (e.g., spatial constraints) can be easily incorporated in the proposed unmixing framework. Future work will consist of reducing the computational complexity of the proposed method.


%


\section*{Acknowledgment}
The authors would like to thank Prof. Jose M. Bioucas-Dias, Dr. Rob Heylen and Dr. Travis R. Meyer for sharing the MATLAB codes of the SUnSAL, AAM and goup/elitist lasso unmixing algorithms.

\ifCLASSOPTIONcaptionsoff
  \newpage
\fi



\bibliographystyle{IEEEtran}
\bibliography{strings_all_ref,ref_all}
%


%




\end{document}